\documentclass[]{spie}  

\newcommand{\ldl}{$\lambda/{\Delta}{\lambda}$}
\newcommand{\microm}{$\mu$m}

\usepackage{amsmath,amsfonts,amssymb, gensymb, deluxetable}
\usepackage{graphicx}
\usepackage[colorlinks=true, allcolors=blue]{hyperref}
\usepackage[labelfont=bf]{caption}

\title{An overview of the NIRSPEC upgrade for the Keck II telescope}

\author[a]{Emily C. Martin}
\author[a]{Michael P. Fitzgerald}
\author[a]{Ian S. McLean}
\author[b]{Gregory Doppmann}
\author[b]{Marc Kassis}
\author[a]{Ted Aliado}
\author[a]{John Canfield}
\author[a]{Chris Johnson}
\author[a]{Evan Kress}
\author[b]{Kyle Lanclos}
\author[a]{Kenneth Magnone}
\author[a]{Ji Man Sohn}
\author[a]{Eric Wang}
\author[a]{Jason Weiss}
\affil[a]{Department of Physics \& Astronomy, University of California, Los Angeles, 430 Portola Plaza, Box 951547, Los Angeles, CA 90095, USA}
\affil[b]{W. M. Keck Observatory, 65-1120 Mamalahoa Highway, Kamuela, HI 96743, USA}

\authorinfo{Further author information: (Send correspondence to E.C.M. or M.P.F.)\\E.C.M.: emartin@astro.ucla.edu; M.P.F.: mpfitz@ucla.edu}

\begin{document} 
\maketitle

\begin{abstract}
NIRSPEC is a 1-5 \microm \ echelle spectrograph in use on the Keck II Telescope since 1999. The spectrograph is capable of both moderate (R=\ldl $\sim$2,000) and high (R$\sim$25,000) resolution observations and has been a workhorse instrument across many astronomical fields, from planetary science to extragalactic observations. In the latter half of 2018, we will upgrade NIRSPEC to improve the sensitivity and stability of the instrument and increase its lifetime. The major components of the upgrade include replacing the spectrometer and slit-viewing camera detectors with Teledyne H2RG arrays and replacing all transputer-based electronics. We present detailed design, testing, and analysis of the upgraded instrument, including the finalized optomechanical design of the new 1-5 \microm \ slit-viewing camera, final alignment and assembly of the science array, electronics systems, and updated software design.  

\end{abstract}

\keywords{Telescopes: Keck, Spectrometers: NIRSPEC, Infrared Instrumentation, Instruments: Upgrades}

\section{INTRODUCTION}
\label{sec:intro}  

\begin{figure} [ht]
\begin{center}
\includegraphics[height=7cm]{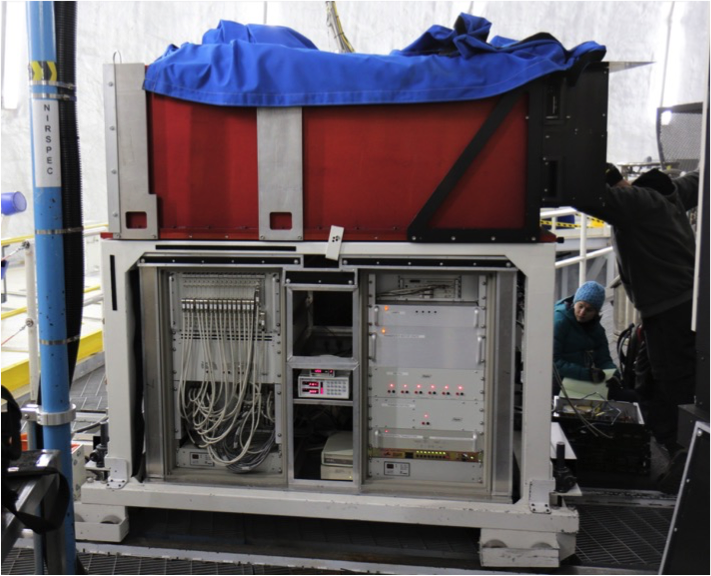}
\end{center}
\caption{ \label{fig:nirspec_nasmyth} 
NIRSPEC on the Keck II Nasmyth Platform. The science dewar (red) sits on top of a frame, holding the transputer-based electronics. The calibration unit (black, upper right) sits in front of the entrance window.}
\end{figure}

NIRSPEC is a cross-dispersed echelle spectrometer for the 1--5 \microm \ wavelength regime deployed on the Nasmyth platform of the Keck II Telescope at W.M. Keck Observatory in Hawaii (Figure~\ref{fig:nirspec_nasmyth}). The instrument features two modes, a moderate resolution (R$\sim$2,000) and a high-resolution (echelle) mode with R$\sim$ 25,000. It also has an independent infrared slit-viewing camera for acquisition and guiding. NIRSPEC was commissioned in 1999 by the IR Lab at UCLA (PI Ian McLean \cite{mclean1998}$^{,}$ \cite{mclean2000}). Since its commissioning, NIRSPEC has proven to be one of the most versatile and useful instruments at Keck. However, infrared array technology has improved significantly over the past 15 years, and the NIRSPEC detectors and electronics are now outdated. We plan to upgrade the instrument to take advantage of improved infrared detector technology, improve the stability and serviceability of the instrument, and extend the useful lifetime of the instrument. 

We will upgrade the current spectrometer (SPEC) detector from an Aladdin III InSb 1024x1024 array to a Teledyne HAWAII-2RG (H2RG) HgCdTe 2048x2048 array, which has smaller pixels and is expected to have lower noise characteristics \cite{martin2014}. Additionally, the current Slit-viewing Camera (SCAM) detector will be upgraded from its current 256x256 PICNIC array (which only operates over 1--2.5 \microm) to a Teledyne H2RG, also optimized for 1--5 \microm. Extending the wavelength coverage of the SCAM will also require a new optical design optimized for the larger bandpass and smaller pixels \cite{martin2016}. We will replace all of the transputer-based readout electronics with the Teledyne SIDECAR ASIC and SIDECAR Acquisition Module (SAM) boards, which are responsible for detector readout and control. Additional upgrades include replacing the entire transputer-based electronics systems, adding thermal and mechanical stability, and upgrading software. Motion control will be handled by Galil motion controllers, and thermal control will be monitored using Lakeshore temperature controllers. 

\begin{figure} [ht]
\begin{center}
\includegraphics[height=7cm]{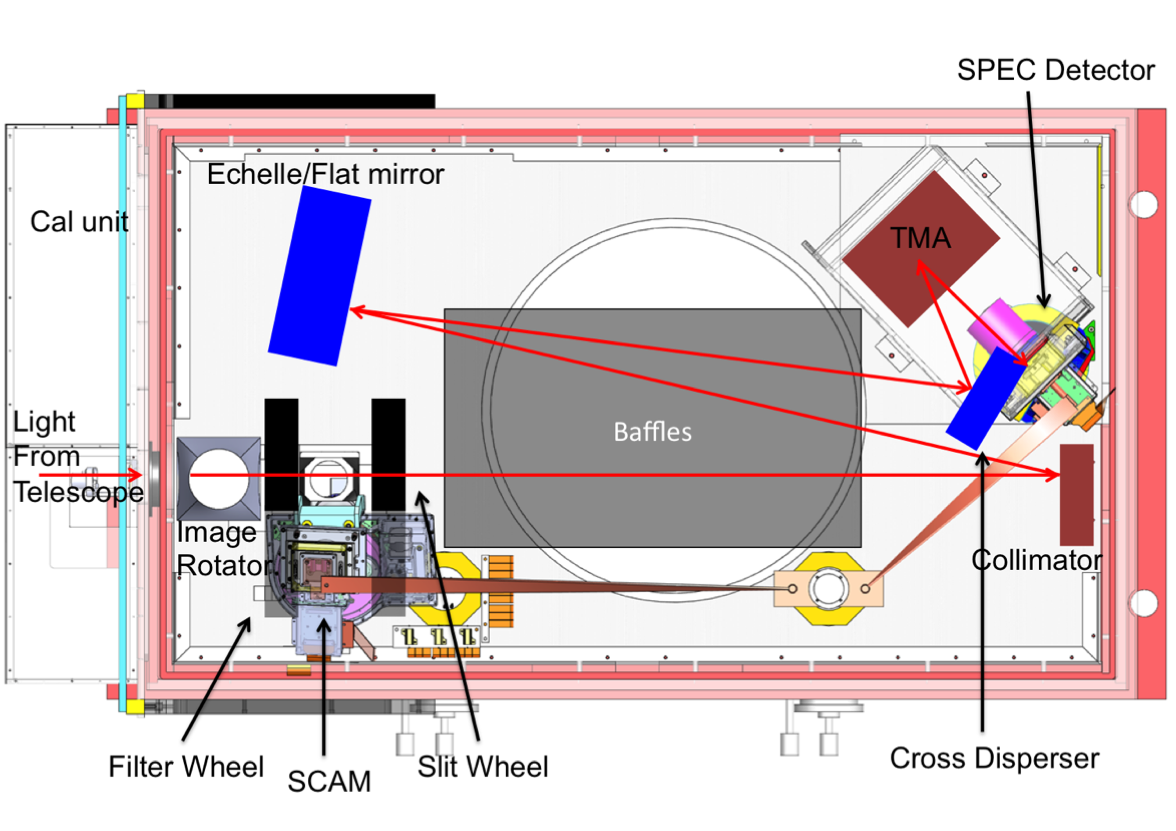}
\end{center}
\caption{ \label{fig:nirspec_solid} 
Top-down solid model of the interior of NIRSPEC. Red arrows trace the light path to the science (SPEC) detector.}
\end{figure} 

Light from the telescope enters NIRSPEC through its entrance window, following the path traced out in Figure~\ref{fig:nirspec_solid}, which shows a top-down schematic of the instrument interior. In the instrument's front-end (prior to the slit), the light is collimated by concave optics in the image rotator arm, passes through an order-sorting filter wheel, and then re-focused into an f/10 beam. Light going through the slit to the back-end is then re-collimated, and either reflected by a flat mirror or dispersed by the echelle grating, depending on the observing mode. The beam is then cross dispersed and imaged onto the detector plane by a three-mirror anastigmat (TMA). Light that doesn't enter the slit is back-reflected into the SCAM, for ease of target alignment. The design is described in detail in McLean et al. (1998) \cite{mclean1998}. 

Both the SPEC and SCAM subsystem upgrades presented mechanical challenges due to the tight mechanical envelopes within the dewar. In particular, the SCAM optical design is constrained by the filter wheel and slit wheel housings on either side, and the focal length of the optical design was constrained by the location of the cold-shield.

In this paper, we provide an overview of the NIRSPEC upgrade and its current status as of June 2018. First, we discuss the upgraded detectors in \S~\ref{sec:detectors}. Then, we present the new optical and opto-mechanical designs for the upgraded SCAM system in \S~\ref{sec:scam}. We discuss stability improvements we are making in \S~\ref{sec:stability}. In \S~\ref{sec:electronics}, we discuss the replacements for the upgraded motion control and housekeeping. We present the software upgrades in \S~\ref{sec:software} and summarize in \S~\ref{sec:summary}.

\section{Detector Upgrades}
\label{sec:detectors}

Table~\ref{tab:detectors} presents an overview of the current and upgraded detector systems. The SPEC and SCAM detectors will both be replaced with Teledyne H2RG arrays. The SPEC array is science grade, with $\sim$ 15 e- CDS readnoise and $\leq$ 0.05 e-/s dark current as measured by Teledyne. The SCAM array is an Engineering A grade detector. There is a significant defect covering $\sim$ 2/3 of the chip, but the SCAM field of view only requires 256x256 pixels. We will only use a sub-array of this detector, allowing us to read out more quickly.

\begin{table}[ht]
\caption{Detector Specifications.} 
\label{tab:detectors}
\begin{center}       
\begin{tabular}{|l|l|l|} 
\hline
\rule[-1ex]{0pt}{3.5ex}   & \textbf{Current System} & \textbf{Upgraded System}  \\
\hline
\rule[-1ex]{0pt}{3.5ex}   & Aladdin III 5-\microm \ cutoff & H2RG 5-\microm \ cutoff   \\
\rule[-1ex]{0pt}{3.5ex}  SPEC & 1024x1024 27-\microm \ pixels & 2048x2048 18-\microm \ pixels  \\
\rule[-1ex]{0pt}{3.5ex}   & 65 e- CDS readnoise & 15 e- CDS readnoise   \\
\hline
\rule[-1ex]{0pt}{3.5ex}   & PICNIC 2.5-\microm \ cutoff & H2RG 5-\microm \ cutoff   \\
\rule[-1ex]{0pt}{3.5ex}  SCAM & 256x256 40-\microm \ pixels & 2048x2048 18-\microm \ pixels  \\

\hline 
\end{tabular}
\end{center}
\end{table}

\subsection{SPEC alignment and assembly}
\label{sec:spec}

\begin{figure} [ht]
\begin{center}
\includegraphics[width=5.8in]{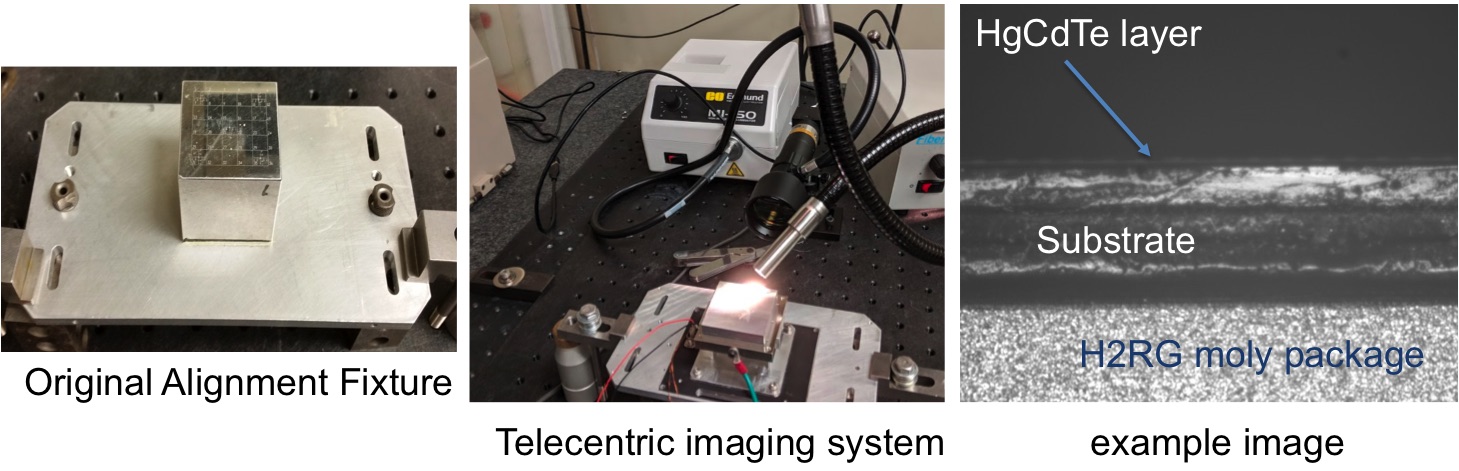}
\end{center}
\caption{ \label{fig:alignment_setup} 
(Left) Alignment fixture used to align the TMA's. (Center) Telecentric imaging system used to measure the height of the HgCdTe layer of the H2RG, relative to the alignment fixture. (Right) Example image of the H2RG taken with the telecentric imaging system.}
\end{figure}

\begin{figure} [ht]
\begin{center}
\includegraphics[width=6in]{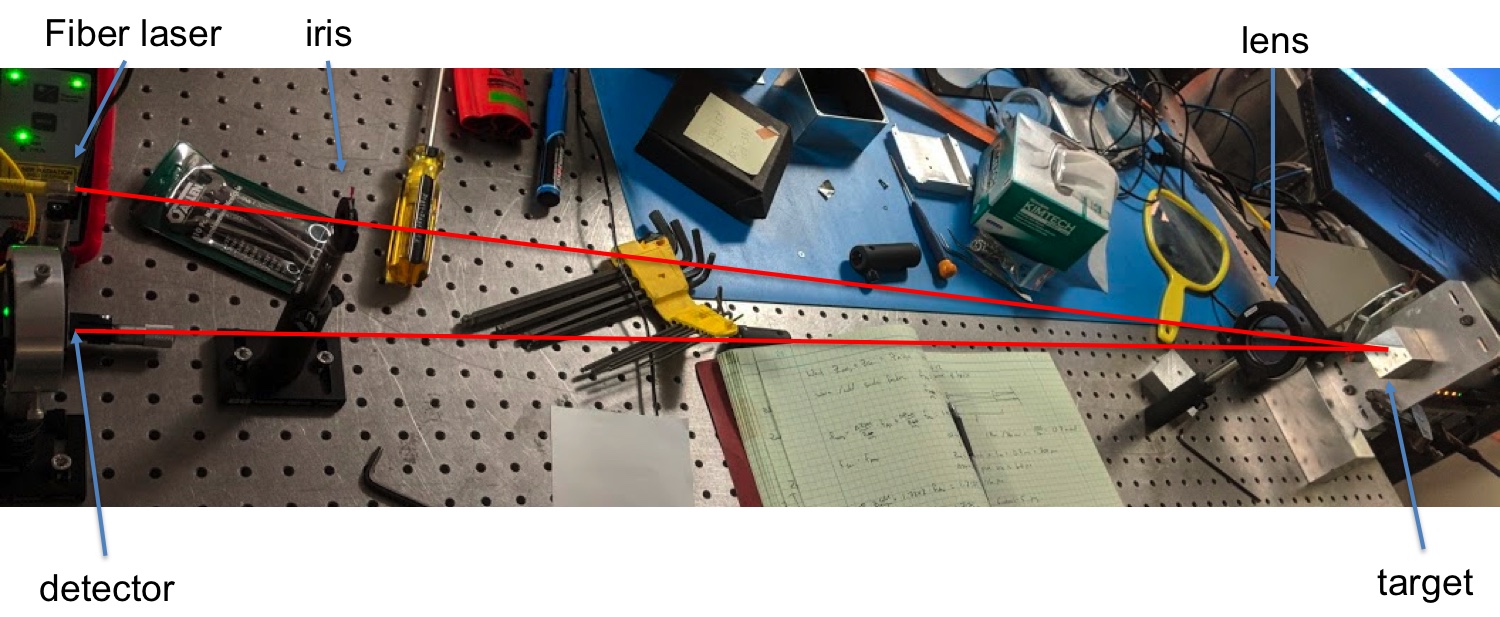}
\end{center}
\caption{ \label{fig:tiptilt} 
Retro-reflection setup to measure tip/tilt offset of the H2RG relative to our alignment fixture. Laser light (left) is collimated by a 1-m focal-length lens and retro-reflected off the surface of the detector, onto a CMOS detector.}
\end{figure}

The replacement SPEC detector head assembly attaches to an alignment bracket in the existing system. We utilized the aluminum alignment fixture that was originally used to align the TMA's to match the focus and tip/tilt of the detector plane with the new H2RG. First, we adjusted the focus using a telecentric imaging system to image the location of the HgCdTe layer relative to the alignment fixture (Figure~\ref{fig:alignment_setup}. Next, we measured tip/tilt by retro-reflecting a laser off of the detector surface (Figure~\ref{fig:tiptilt}). We then manufactured shims to account for both focus and tip/tilt offsets. The shims are installed between the molybdenum block attached to the H2RG and the pedestal it sits on. The pedestal sits on a mounting plate and the entire assembly is then housed within an aluminum enclosure, with a baffled entrance. 

\begin{figure} [ht]
\begin{center}
\includegraphics[width=3in]{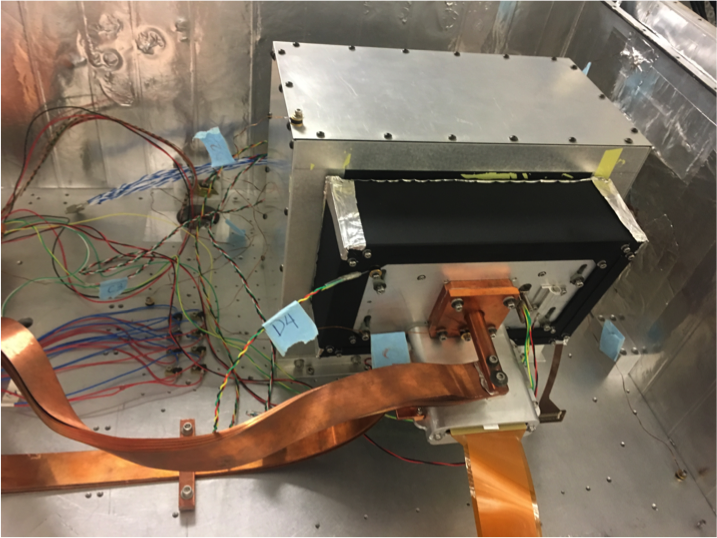}
\end{center}
\caption{ \label{fig:spec_dh} 
Complete prototype spectrometer detector head, assembled for testing.}
\end{figure} 

The fully aligned detector head assembly is shown inside our vacuum cryogenic test chamber in Figure~\ref{fig:spec_dh}. A copper block attached to copper straps provides a direct connection to the 2nd stage cold head of the cryocoolers, cooling the detector to 30 K.

The SIDECAR ASIC is connected via ribbon cable to the H2RG and is housed within an aluminum frame attached at a $90 \degree$ angle relative to the detector plane. A longer ribbon cable attaches to a hermetic feed-through in the dewar wall, which then connects to the SAM card for readout.

\subsection{SCAM detector head assembly}
\label{sec:scam_det}

The SCAM detector head assembly takes advantage of many of the same design features as the SPEC detector head. Significant changes include altering the direction of the ASIC board to extend parallel to the detector and eliminating as much extra height as possible in the detector head enclosure in order to maximize the focal length available to the new optical design.

\section{Slit-viewing Camera Upgrade}
\label{sec:scam}
\subsection{Optical Design}

\begin{figure} [ht]
\begin{center}
\includegraphics[width=3in]{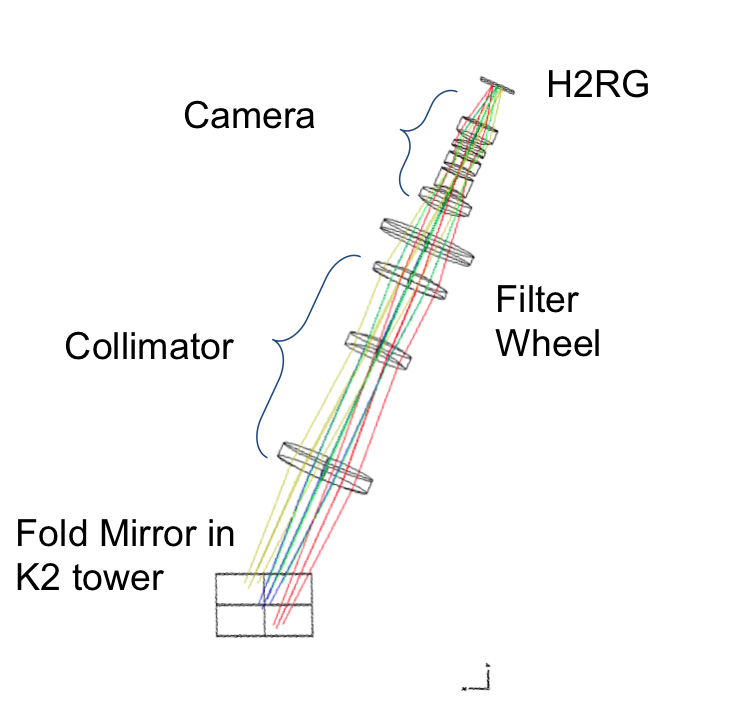}
\includegraphics[width=3in]{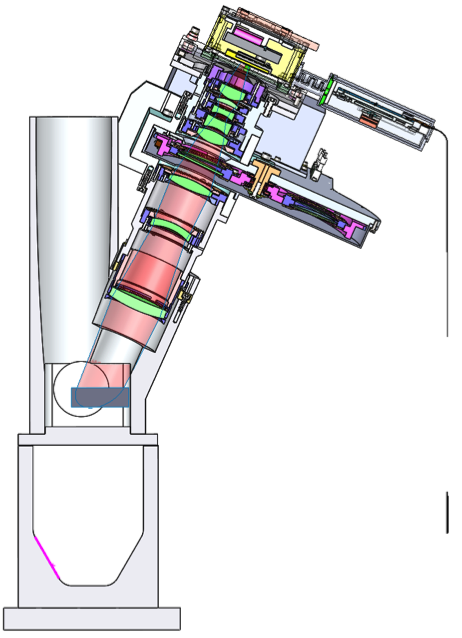}
\end{center}
\caption{ \label{fig:opt_design} 
Left: 3D Layout of the SCAM optical design, from ZEMAX. Right: slice of the SCAM optomechanical solid model.}
\end{figure}

In Martin et al. 2016 \cite{martin2016}, we presented a preliminary optical design for the upgraded Slit-viewing camera. Here, we present the final optical and opto-mechanical design. The upgraded design maintains the same pixel scale as the original SCAM, so with the smaller pixels in the new detector, this required a faster beam of f/2.3, compared to the f/4.6 of the original SCAM. Table~\ref{tab:scam} presents the prescription data for the design and Figure~\ref{fig:opt_design} shows the ZEMAX 3D layout. The final optical design consists of 8 spherical refractive elements and a filter substrate. The three collimating lenses are ZnS, ZnSe, and BaF$_2$, from left to right in the figure. The filter substrate depends on the configuration, with a BK7 shortpass filter for observations in \textit{J, H, K} bands and a Sapphire narrowband filter combined with a Silicon ND3 filter for \textit{L} and \textit{M} band observations. The five camera lenses are BaF$_2$, LiF, ZnSe, ZnS, and BaF$_2$.

\begin{figure} [ht]
\begin{center}
\includegraphics[width=3in]{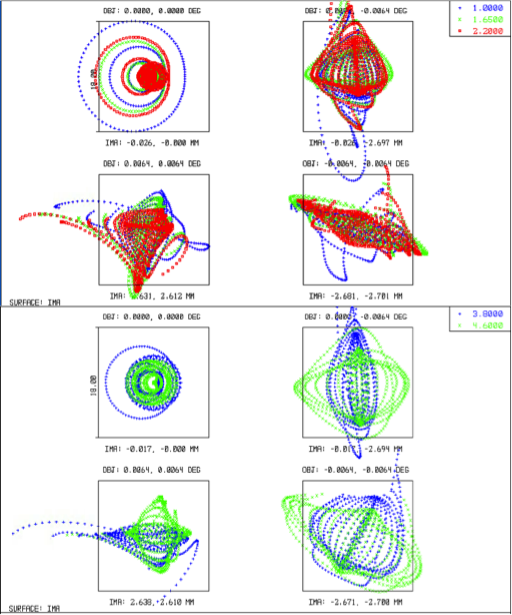}
\includegraphics[width=2.89in]{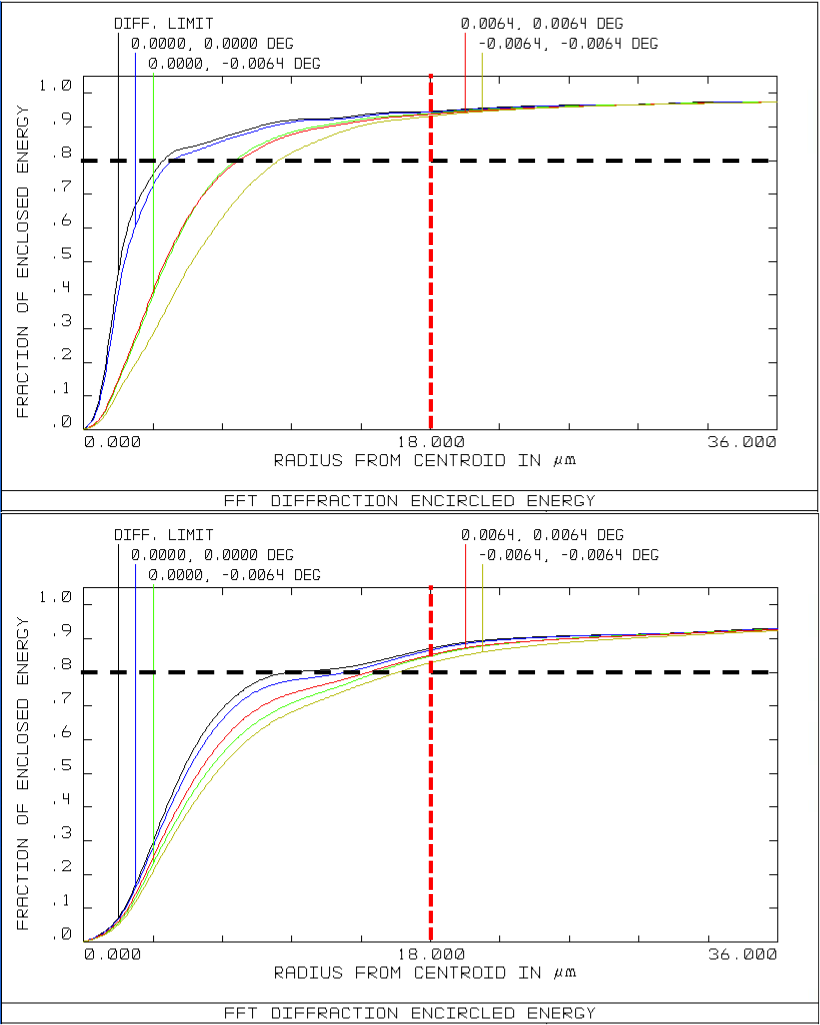}
\end{center}
\caption{ \label{fig:scam_spots} 
Left: Spot diagrams for the two different configurations (\textit{J, H, K}; \textit{L, M}), showing image quality on the SCAM focal plane. Each diagram shows four different spots from four different field locations. The box drawn around each field is the size of a single 18 \microm \ pixel. Right: Encircled energy as a function of radius from the centroid, in \microm. The top panel shows the first configuration, for 1--2.5 \microm, and the bottom panel shows the second configuration, for 3--5 \microm. A black dashed line denotes the 80\% requirement, and a red dashed line marks one pixel. In every field and every wavelength, we meet our requirement.}
\end{figure}

In Figure~\ref{fig:scam_spots} at left, we present spot diagrams of the focal plane for the four fields used for the optimization in ZEMAX. Each color represents a different wavelength. In all four fields and for each wavelength, the design meets our requirement of having the majority of the light fall within one 18 \microm \ pixel. This is further demonstrated in the right panel of Figure~\ref{fig:scam_spots}, where we plot the encircled energy vs. radius for each of the fields. All four fields reach $>$80\% encircled energy within 1 pixel. The image quality in the design is mostly impacted by spherical aberrations in the center of the focal plane. Additional, higher order aberrations are more significant at the edges and corners of the field, though we are still able to maintain the majority of the light within one pixel.

\begin{center}
\begin{deluxetable}{lcccc}
\tabletypesize{\normalsize}
\tablewidth{0in}
\tablecaption{Optical Prescription for the Upgraded Slit-Viewing Camera\label{tab:scam}}
\tablehead{
\colhead{Optic Name} &  
\colhead{Radius of curvature} &    
\colhead{Thickness} &     
\colhead{Material}  &
\colhead{Semi-diameter} \\
\colhead{} &  
\colhead{(mm)} &    
\colhead{(mm)} &     
\colhead{}  &
\colhead{(mm)}
}
\startdata
Fold Mirror & Infinity &  0 & mirror & 23.169\\
Air & --- & 73.858 & Air & --- \\ \hline \\
Collimator 1 &  121.798 &9.348&  ZnS & 24.512\\
---  &  338.142 & 66.609 &  --- & --- \\
Collimator 2 &  $-$41.384 &5.716&  ZnSe & 16.904\\
---  & $-$57.235 & 31.770 &  --- & --- \\ 
Collimator 3 &  103.790 &8.699 & BaF$_2$ & 19.669\\
---  &  $-$69.288 & 15.15 &  --- & --- \\ \hline \\
Filter Config 1\tablenotemark{a} & Infinity & 5.0 & BK7 & 20.0 \\ 
--- & --- & 15.78 & --- & ---  \\ \hline \\
Filter Config 2\tablenotemark{a} & Infinity & 1.0 & Sapphire & 20.0 \\ 
--- & --- & 2.17 & --- & ---  \\ 
--- & Infinity & 1.83 & Silicon & 20.0 \\
--- & --- & 15.78 & --- \\ \hline \\
Camera 1 &  29.828 &10.426 &  BaF$_2$ & 13.466 \\
---  &  $-$44.394 & 3.000 &  --- & --- \\
Camera 2 &  $-$25.373 &8.0& LiF & 8.80\\
---  & 72.427 & 5.000 &  --- & --- \\ 
Camera 3 &  $-$25.366 &4.809 & ZnSe & 8.80\\
---  &  80.438& 4.686 &  --- & --- \\
Camera 4 & $-$73.107 &4.370 & ZnS & 8.80\\
---  &  $-$21.315 & 1.0 &  --- & --- \\
Camera 5 &  19.739 &10.0 & BaF$_2$ & 9.530\\
---  &  Infinity & 22.779 &  --- & --- \\
\enddata
\tablenotetext{a}{There are two configurations for the filters, one for \textit{J}, \textit{H}, \textit{K} observations, and a second for \textit{L} and \textit{M} band observations. The short wavelength configuration only requires a single BK7 filter to block the long wavelength light. The second configuration requires a combination of a narrowband Sapphire filter and an ND3 Silicon filter.}
\end{deluxetable}
\end{center}

\subsection{Opto-mechanical Design}

We show a solid model of the SCAM opto-mechanics from the K2 mirror tower to the SCAM detector in Figure~\ref{fig:opt_design}, on the right. The collimator barrel attaches to the K2 mirror mount in the existing hole pattern there. An angled bracket connects the collimator and camera barrels and also serves to hold the SCAM filter wheel. This design allows for the removal of the SCAM filter wheel while retaining optical alignment of the SCAM collimator and camera. The camera barrel attaches to the detector head via a baffle plate, aligning the image plane on a ``good" portion of the Engineering-grade H2RG. 

The entire SCAM optomechanical design is constructed using Al7075, which has larger tensile strength than the more commonly used Al6061. 

\begin{figure} [ht]
\begin{center}
\includegraphics[width=2.8in]{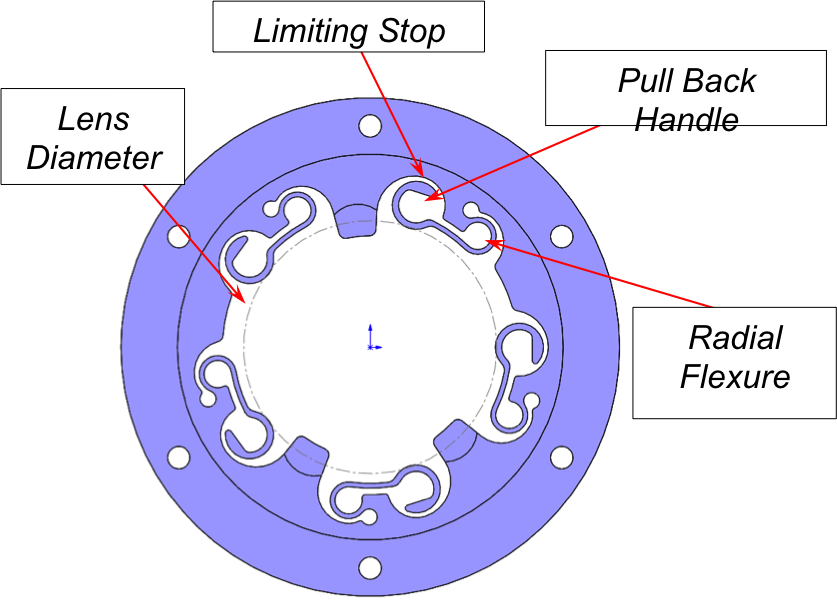}
\includegraphics[width=2.8in]{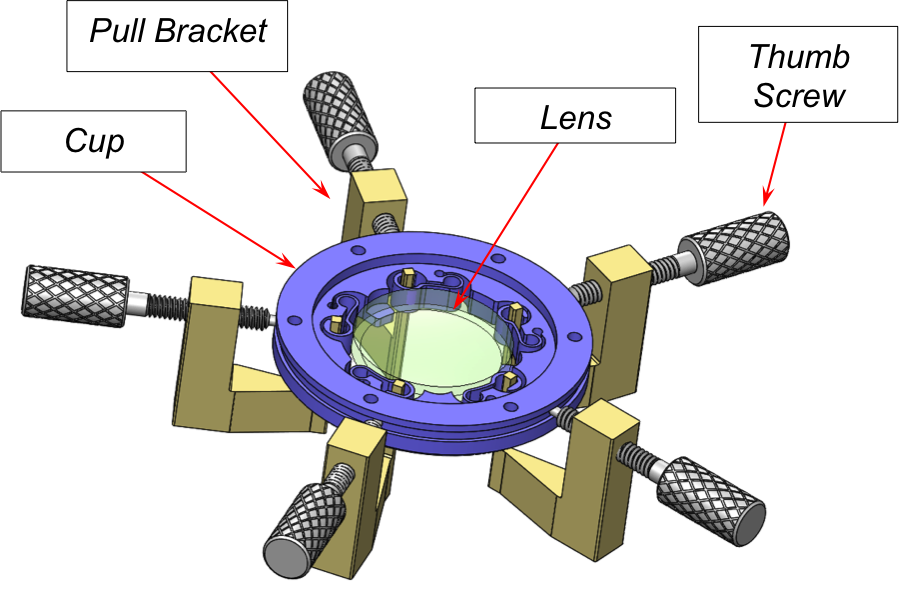}
\end{center}
\caption{ \label{fig:scam_mount} 
The SCAM lens mounts utilize radial flexures to maintain the position of the lens. Thumb screws will be used to hold back the radial flexures while the lens is placed.}
\end{figure} 

\subsubsection{Lens mounts}
The lens mounts in SCAM were inspired by a similar design for the WHIRC instrument's refractive optics \cite{smee2011}. The mounts are designed to be machined using a wire EDM. The radial flexures can be pulled back to the limiting stop using pull brackets with thumb screws, allowing the lens to be placed in the mount (Figure~\ref{fig:scam_mount}).

\begin{figure} [ht]
\begin{center}
\includegraphics[width=2in]{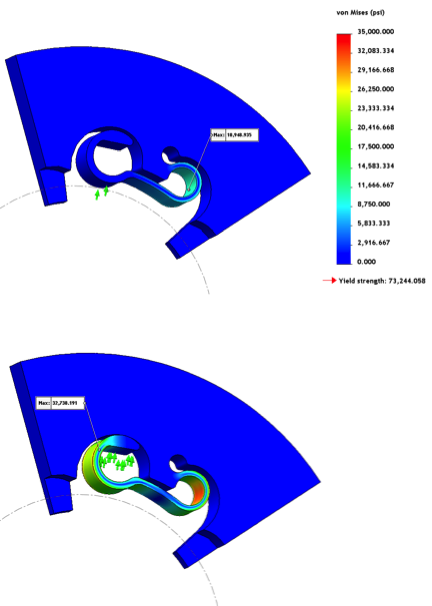}
\end{center}
\caption{ \label{fig:scam_mount_fea} 
Finite Element Analysis of the SCAM lens mounts indicate that the Al7075 is strong enough to meet our expected tensile stresses.}
\end{figure}

A Finite Element Analysis was carried out on the mounts and shows that we can expect the Al7075 to handle the expected stresses, even at maximum flexure, as shown in Figure~\ref{fig:scam_mount_fea}.

\begin{figure} [ht]
\begin{center}
\includegraphics[width=3.5in]{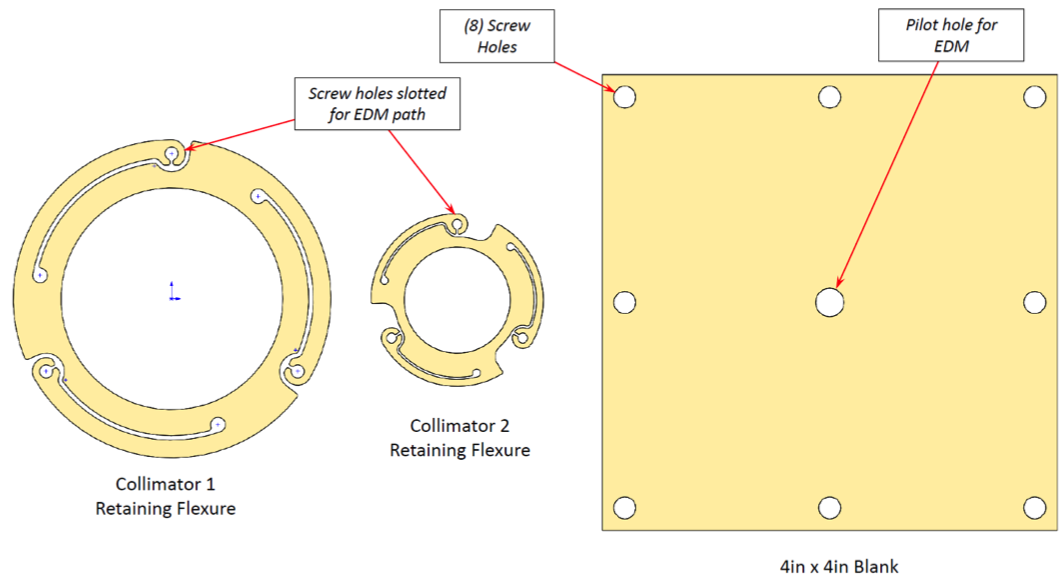}
\end{center}
\caption{ \label{fig:scam_retainer} 
Retaining flexures for the SCAM lenses are machinable with an wire EDM and constructed from Al7075.}
\end{figure}

\begin{figure} [ht]
\begin{center}
\includegraphics[width=5in]{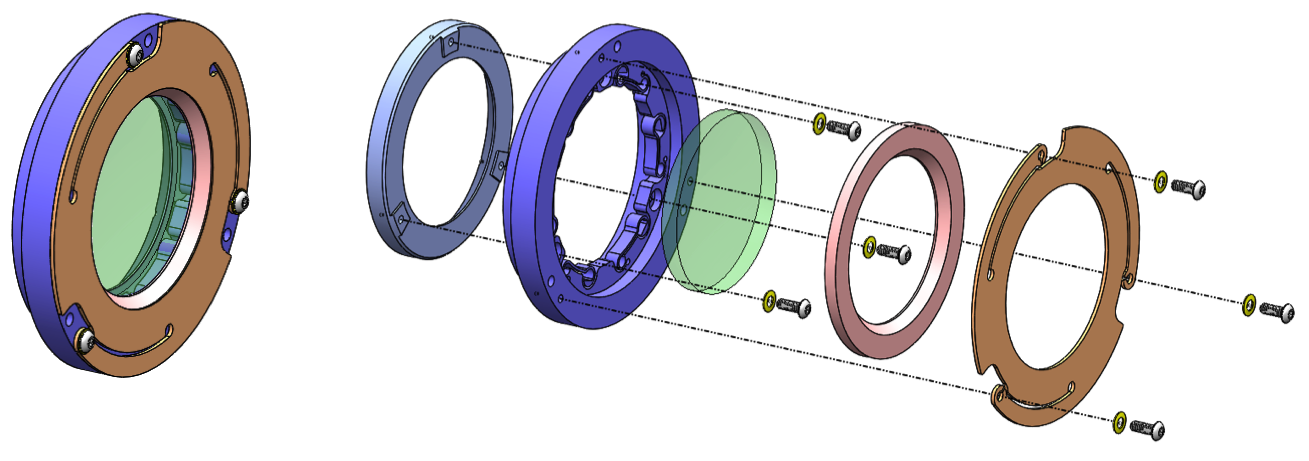}
\end{center}
\caption{ \label{fig:scam_lens_total} 
SCAM lens assembly.}
\end{figure}

Retaining flexures, also constructed from Al7075, hold the mount and retaining rings in place (Figures~\ref{fig:scam_retainer} and \ref{fig:scam_lens_total}). The lens mounts then fit within the custom collimator and camera barrels to maintain optical alignment within our prescribed tolerances. 

\subsubsection{Filter wheel}
\begin{figure} [ht]
\begin{center}
\includegraphics[width=5in]{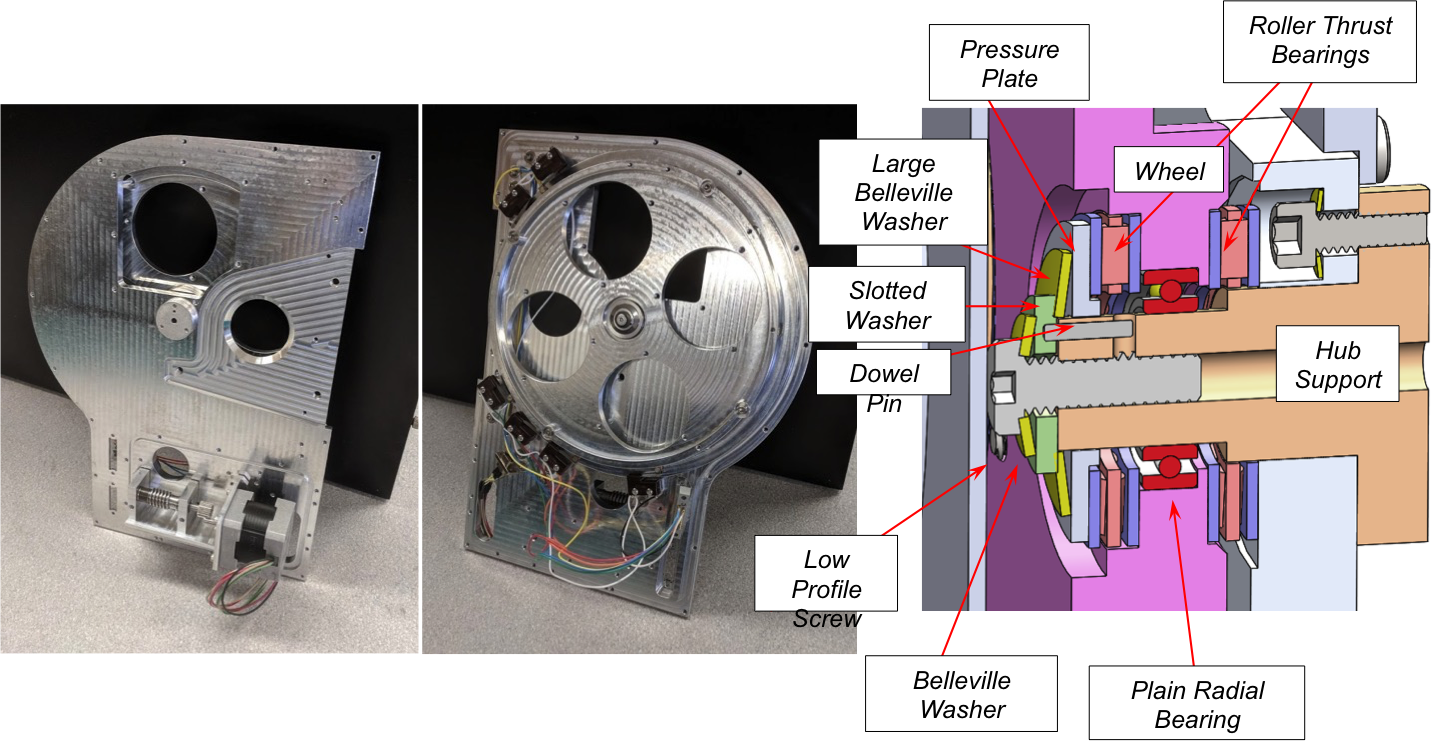}
\end{center}
\caption{ \label{fig:scam_fw} 
SCAM Filter Wheel assembly (left and center), and zoomed-in solid model of the wheel hub.}
\end{figure}

We are adding a filter wheel mechanism to the new SCAM, to ensure that high thermal backgrounds do not saturate the images. The mechanism, shown in Figure~\ref{fig:scam_fw}, contains four filter positions. For observations in the 1--2.5 \microm \ regime, a shortpass filter will block light from wavelengths longer than 2.5 \microm. For \textit{L} and \textit{M} band observations, a combination of a medium-band filter (roughly \textit{L'} and \textit{M'}) with a neutral density (ND3) filter will mitigate the high thermal backgrounds. The wheel also has an open position. The mechanism is driven by a worm gear and will be controlled by a Galil motion controller. 

\section{Stability Improvements}
\label{sec:stability}
We are taking the opportunity of opening NIRSPEC to make several modifications to improve the overall instrumental stability. In Yi et al., 2016 \cite{yi2016}, our team fed a laser frequency comb into NIRSPEC to demonstrate the capabilities of laser combs as wavelength calibration tools. During our testing, we took many hours of images and tracked the internal wavelength solution over time. The instrumental stability of NIRSPEC is most significantly impacted by three things: 1) Echelle grating stability, 2) Cross-disperser grating stability, and 3) Temperature stability of the optical bench. The upgrade allows us to take steps to increase the stability of each of these.

First, we are improving the stability of the echelle grating by adding a mechanical preload to the mechanism. Currently the echelle mechanism is balanced such that any vibrations in the dewar (e.g. by the rotator motor) cause the grating to wobble between gear teeth. By adding a gravitational pre-load, the mechanism will be held against the edge of the tooth, and thus be more stable against instrumental vibrations.

\begin{figure} [ht]
\begin{center}
\includegraphics[width=4in]{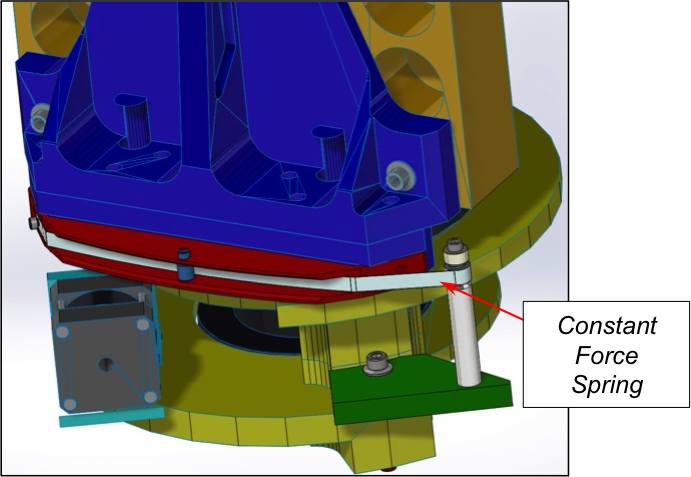}
\end{center}
\caption{ \label{fig:cdg} 
Solid model of the constant force spring that will be added to the cross-disperser grating mechanism to improve its mechanical stability.}
\end{figure}

Secondly, to improve the cross disperser grating mechanism, we are adding on a constant-force spring.  This will pre-load to the mechanism's turret (which rotates horizontally). A solid model of this improvement is shown in Figure~\ref{fig:cdg}.

Finally, we are adding thermal control to the optical bench. Currently, the bench temperature rises and falls with ambient temperatures inside the Keck II dome. The cryo-coolers are run at full speed and there is no heater control on the bench to maintain a more constant temperature. We will add two heaters connected to a Lakeshore 336 controller to maintain temperature stability to $\sim$0.1 K, whereas current temperatures vary at $\sim$ few K per day. 

\section{Electronics Upgrade}
\label{sec:electronics}
All transputer-based electronics will be replaced with modern counterparts, all of which will be compatible with the Keck Task Library (KTL) protocol. Detector readout and control will be managed by Teledyne SIDECAR ASIC and SAM cards, as discussed above. Motion control will be managed by two 8-axis Galil motion controllers, which will drive all of the existing mechanisms (6 cryogenic, 3 in calibration unit) and the new SCAM filter wheel. Arc and flat lamps will also be powered by the Galil. It was necessary to purchase an additional optoisolator card for the Galil to use in conjunction with the extended I/O ports. Housekeeping will be handled by new Lakeshore temperature controllers and the existing Varian pressure gauge. We will use a Lakeshore 218 for general temperature monitoring within the dewar, and two Lakeshore 336 units for detector and optical bench temperature control. Four Eaton power controllers will be able to control all components individually.

All electronics will fit within the original 19" electronics racks located underneath the dewar on the instrument cart. The electronics are housed in standard 19" enclosures, with environmental cabling connecting each component to the dewar or cal unit. Cabling will be centralized and rerouted through the back of the racks to allow for easier servicing of the electronics components.

\section{Software Upgrade}
\label{sec:software}
Alongside electronics hardware upgrades, we are upgrading several components of the NIRSPEC software. The previous transputer-based software must be replaced, allowing us to modernize the software. All computer systems will be Linux servers, and all software will utilize KTL keywords for interprocess communications.

Detectors will be run by a GigE interface version of the Sidecar software from Teledyne, allowing all detector-related control to run on Linux, instead of the previous generation which required Windows control computers. Motion control will utilize galildisp software that has previously been implemented for other instruments at Keck Observatory. Housekeeping interfaces are simple, serial-like, and will be easily controllable with the Linux host computer. 

Additionally, upgrades will be made to the graphical user interfaces (GUIs) that the observer can use to control the instrument. Motion control GUIs will be written in Python using PyQt. Detector control will use Ginga for FITS image display and detector controls. In all cases, the software upgrades are relying heavily on heritage work done for existing instruments at Keck or at other observatories (e.g. Subaru).

\section{Summary}
\label{sec:summary}

We presented an overview of the NIRSPEC upgrade for the Keck II telescope, which will be installed in Fall 2018. The major components of the upgrade include replacing both the spectrometer and slit-viewing camera detectors with Teledyne H2RG arrays, replacing the entire slit-viewing camera so that it will be operable across 1--5 \microm, replacing all transputer-based electronics, adding instrumental stability, and improving the software. The upgraded components are being manufactured in the UCLA IR Lab and will be installed in the NIRSPEC dewar on the summit of MaunaKea. The upgrade is expected to improve the efficiency and capabilities of NIRSPEC. The increased detector sensitivity will allow observers to observe objects up to $\sim$ 1 mag fainter in the same exposure time, when limited by detector noise.

\acknowledgments 
We would like to thank Teledyne Imaging Sensors for their gracious support of the UCLA IR Lab throughout the prototyping and development process. We also thank W. M. Keck Observatory, which is operated as a scientific partnership among the California Institute of Technology, the University of California and the National Aeronautics and Space Administration. The Observatory was made possible by the generous financial support of the W.M. Keck Foundation. The authors wish to recognize and acknowledge the very significant cultural role and reverence that the summit of Maunakea has always had within the indigenous Hawaiian community.  We are most fortunate to have the opportunity to conduct observations from this mountain.

We would like to thank Michael Thacher for his generous contributions to help enable this work.

We would also like to thank the Heising Simons Foundation for their generous support of this project.

This work was made possible in part by the National Science Foundation award AST-1532315.

\bibliography{spie2018bib} 
\bibliographystyle{spiebib} 

\end{document}